\documentclass[fleqn,usenatbib]{mnras}

\usepackage{newtxtext,newtxmath}
\usepackage[T1]{fontenc}

\usepackage{graphicx}	
\usepackage{amsmath}	
\usepackage{amssymb}	
\usepackage{capt-of}

\title[Keeping M-Earths Habitable]{Keeping M-Earths Habitable in the Face of Atmospheric Loss by Sequestering Water in the Mantle}

\author[Moore \& Cowan]{
Keavin Moore,$^{1,2}$\thanks{E-mail: keavin.moore@mail.mcgill.ca}
Nicolas B. Cowan$^{1,2,3}$
\\
$^{1}$Department of Earth \& Planetary Sciences, McGill University, 3450 rue University, Montr\'{e}al, QC H3A 0E8, Canada\\
$^{2}$McGill Space Institute, McGill University, 3550 rue University, Montr\'{e}al, QC H3A 2A7, Canada\\
$^{3}$Department of Physics, McGill University, 3600 rue University, Montr\'{e}al, QC H3A 2T8, Canada
}

\date{Accepted XXX. Received YYY; in original form ZZZ}

\pubyear{2020}

\begin{document}
\label{firstpage}
\pagerange{\pageref{firstpage}--\pageref{lastpage}}
\maketitle

\begin{abstract}
Water cycling between Earth's mantle and surface has previously been modelled and extrapolated to rocky exoplanets, but these studies neglected the host star. M-dwarf stars are more common than Sun-like stars and at least as likely to host temperate rocky planets (M-Earths). However, M dwarfs are active throughout their lifetimes; specifically, X-ray and extreme ultraviolet (XUV) radiation during their early evolution can cause rapid atmospheric loss on orbiting planets. The increased bolometric flux reaching M-Earths leads to warmer, moister upper atmospheres, while XUV radiation can photodissociate water molecules and drive hydrogen and oxygen escape to space. Here, we present a coupled model of deep-water cycling and water loss to space on M-Earths to explore whether these planets can remain habitable despite their volatile evolution. We use a cycling parameterization accounting for the dependence of mantle degassing on seafloor pressure, the dependence of regassing on mantle temperature, and the effect of water on mantle viscosity and thermal evolution.  We assume the M dwarf's XUV radiation decreases exponentially with time, and energy-limited water loss with 30\% efficiency. We explore the effects of cycling and loss to space on planetary water inventories and water partitioning. Planet surfaces desiccated by loss can be rehydrated, provided there is sufficient water sequestered in the mantle to degas once loss rates diminish at later times. For a given water loss rate, the key parameter is the mantle overturn timescale at early times: if the mantle overturn timescale is longer than the loss timescale, then the planet is likely to keep some of its water.
\end{abstract}

\begin{keywords}
planets and satellites: atmospheres -- planets and satellites: interiors -- planets and satellites: tectonics -- planets and satellites: terrestrial planets -- planets and satellites: oceans -- stars: low-mass
\end{keywords}

\section{Introduction}\label{sec:intro}

Habitability critically depends on the presence of liquid water on the surface of a planet. Earth is the only planet in the Universe with confirmed surface oceans and life as we know it, and as such, it is our template for a habitable planet \citep{langmuir12}. The habitable zone (HZ) around a star is usually defined by the stellar flux at a given orbital distance, which influences the surface temperature \citep{kasting93} --- planets at the hot inner edge vaporize their oceans, while the oceans of a planet at the cold outer edge will freeze. Water is not only important for biological processes; it also influences planetary climate through the silicate weathering thermostat \citep{walker81, sleep01, abbot12, alibert14}.

It must be noted that the presence of liquid surface water may constitute habitability in the classical definition \citep{kasting93}, but this does not necessarily mean the planet is hospitable and conducive to the origin of life. Rather, liquid surface water is a first step towards life as we know it, and our results only support the presence of liquid surface water and thus the classical definition of habitability. Recent studies find that the orbital environment around M dwarfs may lack the necessary levels of UV radiation to form RNA monomers, which are critical to the development of Earth-like biology \citep{ranjan17, rimmer18}; however, the scarcity of UV may be overcome by transient flaring events. We must be careful in assuming liquid surface water means life due to the multitude of factors that allowed life to originate on the early Earth.

\subsection{Water Cycling}

It is speculated that there is at least as much water sequestered in the mantle as is present on the surface of Earth \citep{hirschmann06}. Current estimates based on experimental data put the water capacity of Earth's mantle at 12 terrestrial oceans (TO, where 1 TO $\approx 1.4 \times 10^{21}$ kg; \citealt{hauri06}; \citealt{cowan14}). 

Water is exchanged between the surface and mantle reservoirs of Earth on geological timescales. This water cycle is mediated by plate tectonics, which depend not only on the cool, brittle lithospheric plates, but also on the viscous, flowing mantle (see, e.g., \citealt{hirschmann06}; \citealt{langmuir12}). Water dissolved in the mantle decreases its viscosity and allows it to flow more readily, a requirement for plate tectonics \citep{hauri06}. As plates separate from one another at mid-ocean ridges, the mantle below ascends to fill the gap and melts due to depressurization, and volatiles are released into the ocean by degassing as the new ocean crust solidifies. The ocean crust then spreads away towards subduction zones, and its minerals become hydrated due to hydrothermal interactions with seawater. This volatile-rich slab is then subducted back into the mantle. While most of the volatiles will contribute to water-rich magmas at convergent margin volcanoes, some water will continue into the deep mantle, regassing water into the interior.

\subsection{Water Loss to Space}

Roughly 70\% of stars in the Galaxy are M dwarfs \citep{henry04}, and about 30\% are host to at least one rocky exoplanet in the HZ \citep{dressing15}. We should then expect that 90-99\% of temperate terrestrial planets orbit an M dwarf instead of a Sun-like star; henceforth, we call such planets M-Earths.

M dwarfs are more active than Sun-like stars \citep{scalo07}, specifically in the X-ray and extreme ultraviolet, collectively known as XUV. An M dwarf emits more XUV radiation during the first billion years of its lifetime, as the star evolves onto the main sequence. Young planets orbiting within the HZ may lose multiple oceans of water to space as they are bombarded by XUV radiation. Water molecules are photodissociated high in the planetary atmosphere \citep{wordsworth13, wordsworth14, luger15, bolmont16, schaefer16, wordsworth18, fleming20}; the lighter hydrogen is lost to space, while the heavier oxygen remains behind, either reacting with the surface or accumulating in the atmosphere and creating a biosignature false positive (see, e.g., \citealt{wordsworth13, wordsworth14, wordsworth18}). Some oxygen may also hydrodynamically escape, dragged to space by the escaping hydrogen (e.g., \citealt{hamano13, luger15}). Studies also indicate that factors of 5--10 more XUV irradiation than modern-day Earth can lead to runaway atmospheric loss \citep{tian08}, and that several Gyr-old planet-hosting M dwarfs may output 5--100 times more XUV radiation than the Sun today (e.g., \citealt{ribas16, ribas17}; \citealt{youngblood16}). Moreover, it has recently been indicated that M dwarfs remain more active in the extreme-UV (EUV) for a given age than solar-type stars (see, e.g., Fig. 6 of \citealt{france18}). 

We hypothesize that a planet whose surface becomes desiccated by loss of water to space can recover an ocean through the degassing of water sequestered in the mantle. This will depend on the initial amount of water partitioned between the surface and mantle, the mantle overturn timescale, and the XUV-driven water loss rate and timescale. While the deep-water cycle and atmospheric loss have been separately modelled in previous work, we seek to couple these two phenomena, combining aspects of geophysics, astronomy, and space physics.

The paper continues as follows. We describe the cycling and loss equations of our model in Section~\ref{sec:equations}, present our cycling results for various initial water inventories and loss rates in Section~\ref{sec:results}, and discuss the results of our study in Section~\ref{sec:discussion}.

\section{Water Cycling \& Loss Model} \label{sec:equations}

\subsection{Previous Work}

The deep-water cycle of Earth has previously been represented using two-box models. These models account for regassing of water from surface to mantle through subduction of hydrated basaltic oceanic crust, and degassing from mantle to surface by mid-ocean ridge volcanism. \citet{mcgovern89} incorporated reduction of mantle viscosity through the addition of regassed water, while parameterizing degassing and regassing rates as dependent on the amount of volatiles present in the mantle and basaltic oceanic crust, respectively, along with mid-ocean ridge spreading rate and subduction efficiency. 

The mantle-temperature-dependent model of \citet{schaefer15}, based on the model of \citet{sandu11}, also included mantle viscosity and two convection regimes: single layer and boundary layer. \citet{komacek16} simplified the mantle-temperature-dependent model of \citet{schaefer15}, and replaced the degassing rate with the seafloor-pressure-dependent degassing parameterization of \citet{cowan14} to create a hybrid model. 

There are various water loss rates throughout the literature; for example, \citet{wordsworth14} note that an N\textsubscript{2}-poor planet could lose up to 0.07 TO/Gyr, while loss rates from \citet{luger15} range from 0.02 TO/Gyr to about 2 TO/Gyr, depending on initial water inventory and orbital distance within the HZ.

\subsection{Cycling \& Loss Equations}

\begin{figure}
\centering
\includegraphics[width=0.45\textwidth]{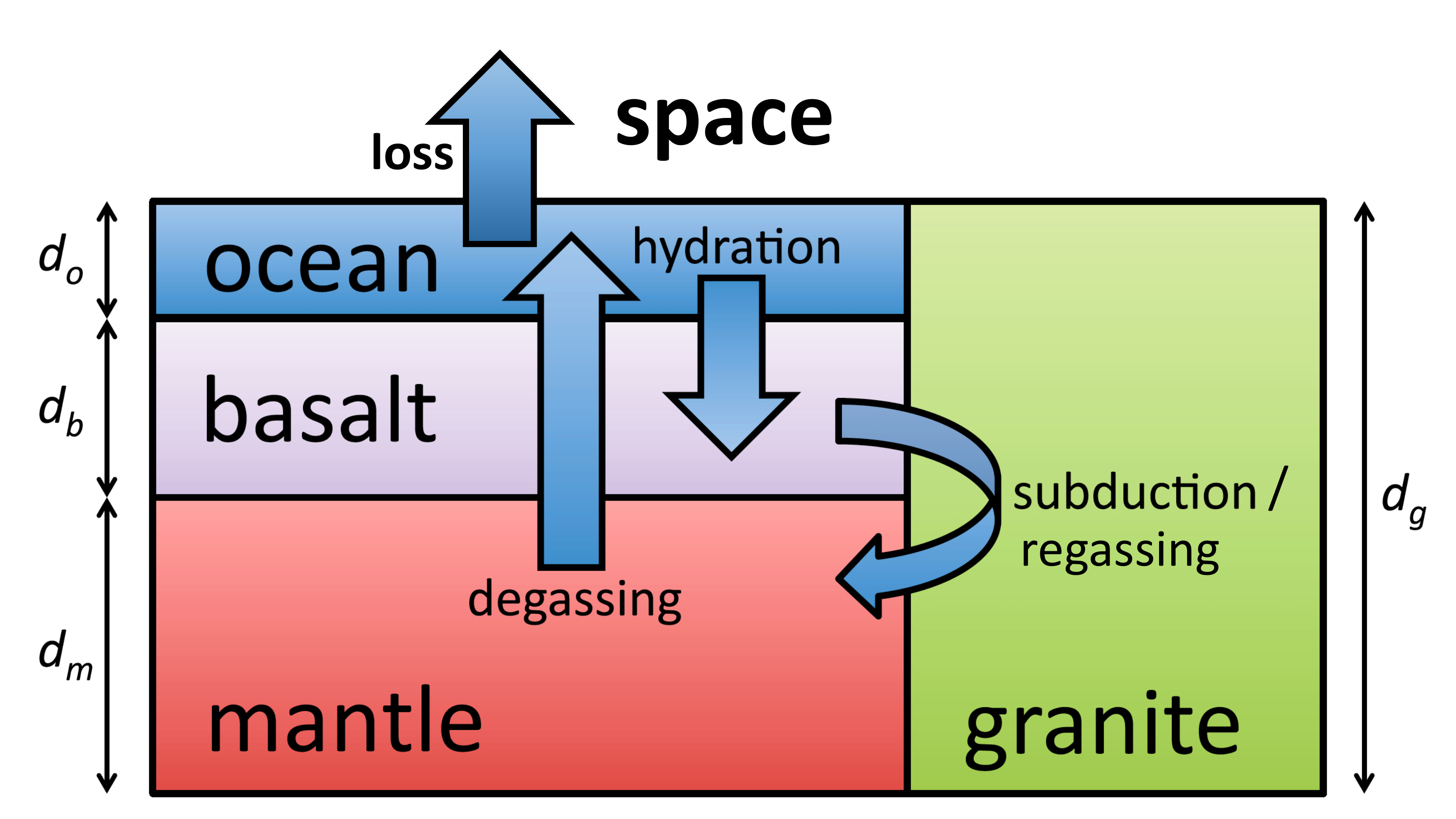}
\caption{Two-box model of water cycling between surface and mantle reservoirs on Earth, adapted from \citet{cowan14} to include water loss to space (bolded). Water is degassed from the mantle to the surface through mid-ocean ridge volcanism, and regassed from the surface to the mantle through subduction of hydrated basaltic oceanic crust. Water is lost to space directly from the surface reservoir for simplicity, and is driven by XUV radiation from the host M dwarf, which decreases exponentially with time \citep{luger15}.}
\label{fig:twoboxsink}
\end{figure}

We use the time-dependent hybrid cycling model of \citet{komacek16} and parameterize the water loss of \citet{luger15} to represent the cycling and loss to space of water on an M-Earth. Our model accounts for the fact that hydration depth of ocean crust is likely affected by mantle temperature, $T$, more than seafloor pressure, $P$, and that degassing would shut off at late times when the mantle is cool. Meanwhile, it has been shown that degassing should be $P$-dependent \citep{kite09}. Our two-box $+$ sink model is shown schematically in Fig.~\ref{fig:twoboxsink}, including surface and mantle reservoirs, exchange between the two, and water loss to space directly from the surface reservoir, for simplicity. 

Any changes or additions to the relevant thermal evolution and cycling equations from \citet{komacek16} are described here and in Appendices \ref{sec:thermaleqns} and \ref{sec:improvements}. The thermal evolution and cycling equations were non-dimensionalized by \citet{komacek16} to emphasize the physical processes over the control variables themselves. While we use the non-dimensionalized code for our simulations, we present the dimensionful equations here. Appendix \ref{sec:params} contains a cheat sheet of all the model variables and parameters.

The model developed by \citet{komacek16} incorporates $P$-dependent degassing \citep{cowan14} and $T$-dependent regassing \citep{schaefer15}. The authors note that this hybrid model may be the most realistic deep-water cycling model of their study; for this reason, we use it as our representative water cycling parameterization.

We restore the piecewise degassing limit of \citet{cowan14}; while this makes the equation harder to manipulate analytically, it ensures that a parcel of mantle cannot degas more water than it contains. The change in mantle water mass $W_{\mathrm{m}}$ with time $t$ is,
\begin{equation}\label{eqn:dx/dt_hyb}
\begin{split}
    \frac{d W_{\mathrm{m}}}{dt} & = L_{\mathrm{MOR}} S(T) \Biggl[x_{\mathrm{h}} \rho_{\mathrm{c}} \chi_{\mathrm{r}} d_{\mathrm{h}}(T) \\
    & - x \rho_{\mathrm{m}} d_{\mathrm{melt}} \min \biggl[ f_{\mathrm{degas},\oplus} \left(\frac{P}{P_\oplus} \right)^{-1}, ~1 \biggr] \Biggr],
\end{split}
\end{equation}
where the first term on the right-hand side is the regassing rate, $w_{\downarrow}$, and the second is the degassing rate, $w_{\uparrow}$. The other variables are as follows: $L_{\mathrm{MOR}}=3 \pi R_\mathrm{p}$ is the mid-ocean ridge length (with $R_{\mathrm{p}}$ the planetary radius), and $S(T)$ is the $T$-dependent spreading rate. The mass fraction of water in the hydrated crust is $x_{\mathrm{h}}$, $\rho_{\mathrm{c}}$ is the density of the crust, $\chi_{\mathrm{r}}$ is the subduction efficiency, and the hydrated layer depth is a function of mantle temperature, $d_{\mathrm{h}}(T)$. The mantle water mass fraction is $x$. The density of the upper mantle is $\rho_{\mathrm{m}}$, $d_{\mathrm{melt}}$ is the mid-ocean ridge melting depth, $f_{\mathrm{degas},\oplus} = 0.9$ is the nominal value of melt degassing for present-day Earth, $P$ is the seafloor pressure, and $P_\oplus$ is the seafloor pressure of Earth. The pressure dependence is defined as a power law, using the nominal value from \citet{cowan14} for the exponent.

The equivalent cycling equation for the surface water mass $W_{\mathrm{s}}$ is,
\begin{equation}\label{eqn:ds/dt_hyb}
\begin{split}
    \frac{d W_{\mathrm{s}}}{dt} & = L_{\mathrm{MOR}} S(T) \Biggl[x \rho_{\mathrm{m}} d_{\mathrm{melt}} \min \biggl[ f_{\mathrm{degas},\oplus} \left(\frac{P}{P_\oplus} \right)^{-1}, ~1 \biggr] \\
    & - x_{\mathrm{h}} \rho_{\mathrm{c}} \chi_{\mathrm{r}} d_{\mathrm{h}}(T) \Biggr] - \min \left[\phi_{\mathrm{loss}} \exp \left(\frac{-t}{\tau_{\mathrm{loss}}}\right),~ \frac{W_{\mathrm{s}}}{\tau_{\mathrm{step}}} \right].
\end{split}
\end{equation}
Here, the first term on the right-hand side is now the degassing rate, $w_{\uparrow}$, and the second is the regassing rate, $w_{\downarrow}$. The third term on the right-hand side is the water loss rate, $w_{\mathrm{loss}}$, a decreasing exponential based on the M dwarf XUV evolution with time in Fig.~1 of \citet{luger15}, which utilizes the stellar models of \citet{ribas05}; note that we assume the XUV radiation simply decreases exponentially from its initial value, with the loss of water to space linearly correlated to this evolution. Note also we do not directly model the stellar evolution, nor do we account for a planetary magnetic field; as a result, we do not include flare- or stellar-wind-driven loss in this study. Recent studies support stellar-wind-driven ion pick-up escape leading to rapid, complete atmospheric erosion for planets orbiting `old' (i.e., several Gyr) M dwarfs like Proxima Centauri b \citep{airapetian17, garcia17}, in the absence of a source of replenishment. Although ion escape is likely important for M-Earths, we do not include it in our current study. Instead, we solely focus on the energy-limited escape of \citet{luger15}, adopting the same efficiency of 30\% to test similar loss rates.

Our loss parameterization is piecewise-defined so that we do not lose more water than present on the surface in a given timestep. The exponential definition of loss to space stems from the exponential decrease of the M dwarf's XUV luminosity, and includes a loss factor, $\phi_{\mathrm{loss}}$, and loss timescale, $\tau_{\mathrm{loss}}$. The former accounts for the range of water loss rates in the literature, and represents the energy-limited loss rate in a single variable, $\phi_{\mathrm{loss}}$; the latter represents the e-folding timescale of water loss to space, i.e., water loss is reduced by $1/e$ after $\tau_{\mathrm{loss}}$. For simplicity, we model loss of water directly to space from the surface. This approximation should be valid if the atmosphere is hot -- and hence moist -- in the era of high XUV. We include the hydrated layer check from \citet{schaefer15} to ensure that the hydrated layer holds no more water than the surface itself. 

The model explicitly depends on mantle temperature via the mid-ocean ridge spreading rate, $S(T)$. Moreover, we stop degassing if the mantle cools below the solidus temperature, since no more melt will be present in the boundary layer. We calculate the wet solidus temperature using the parameterization of \citet{katz03}, since water in the mantle depresses the solidus of silicate minerals.

\section{Simulation Results}\label{sec:results}

We run simulations for various initial total water inventory, $W_{\mathrm{m},0} + W_{\mathrm{s},0}$,  loss factor, $\phi_{\mathrm{loss}}$, and loss timescale, $\tau_{\mathrm{loss}}$. All simulations are run for 15 Gyr to allow our model to reach a steady state, if possible. Our parameter exploration is shown in Table \ref{tab:param_space}. We test four orders of magnitude for both $\phi_{\mathrm{loss}}$ and $\tau_{\mathrm{loss}}$, due to the range of loss rates in the literature, and because of the large uncertainties in observations and models of M dwarfs. We also test various initial water inventories, since planets are expected to form with different volatile inventories due to stochastic delivery and accretion \citep{raymond04, raymond09}. Note that all simulations begin with an initial mantle temperature of $T_0 = 3200$ K, i.e., $T_0 = 2 T_{\mathrm{ref}}$, where $T_{\mathrm{ref}}$ is the reference temperature used in our thermal evolution calculations (detailed in Appendix \ref{sec:thermaleqns}).

\begin{table}
    \centering
    \begin{tabular}{c|c|c}
    \hline
        Name & Parameter & Values Tested  \\
    \hline
         Total water mass & $W_{\mathrm{m},0} + W_{\mathrm{s},0}$ [TO] & 0.1, 1, 10, 25 \\
         Mantle temperature & $T_0$ [K] & 3200 \\
         Loss factor & $\phi_{\mathrm{loss}}$ [TO/Gyr] & 0.1, 1, 10, 100 \\
         Loss timescale & $\tau_{\mathrm{loss}}$ [Gyr] & $10^{-3}$, $10^{-2}$, $10^{-1}$, $1$ \\
    \hline
    \end{tabular}
    \caption{Parameter space for initial total water inventory, $W_{\mathrm{m},0} + W_{\mathrm{s},0}$, loss factor, $\phi_{\mathrm{loss}}$, and loss timescale, $\tau_{\mathrm{loss}}$. We initiate all simulations with the same mantle temperature, $T_0 = 2 T_{\mathrm{ref}}.$ Water mass is expressed in units of terrestrial oceans, where 1 TO $\approx 1.4 \times 10^{21}$ kg.}
    \label{tab:param_space}
\end{table}

\begin{figure}
\centering
\includegraphics[width=0.45\textwidth]{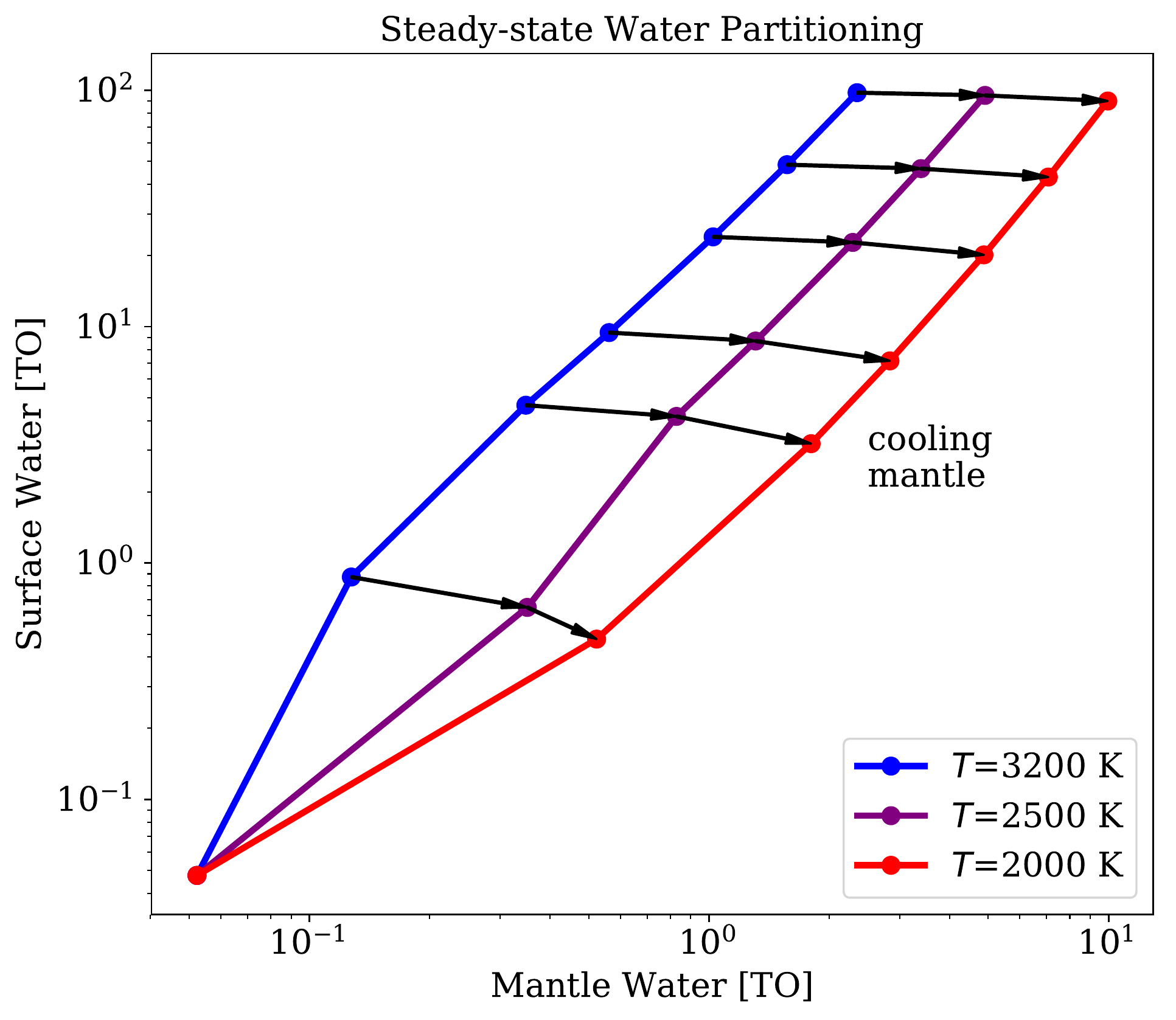}
\caption{\label{fig:steadystate_temps} Steady-state water partitioning between surface, $W_{\mathrm{s}}$, and mantle, $W_{\mathrm{m}}$, reservoirs in units of terrestrial oceans (TO), for different mantle temperatures. Since the mantle cools with time, we would expect the steady state to change as well. Indeed, the steady-state curves shift towards the lower right; cooler temperatures lead to less surface water and more mantle water. For a given total water inventory and mantle temperature, there is a unique steady-state partitioning of water, and that is precisely the partitioning we use when initializing simulations in Figs.~\ref{fig:hyb_exploss} \& \ref{fig:Wloss_tloss_water}.}
\end{figure}

Each of the total water inventories from Table \ref{tab:param_space} is first run to steady-state partitioning between the mantle and surface reservoirs (Fig.~\ref{fig:steadystate_temps}), without loss to space and at a constant mantle temperature of $T=3200$ K. To visualize the evolution of steady-state water partitioning as the mantle cools with time, we run simulations for various initial water inventories ($W_{\mathrm{m},0}+W_{\mathrm{s},0} = $ 0.1, 1, 5, 10, 25, 50, 100 TO) at three constant mantle temperatures, $T = $ 3200 K, 2500 K, 2000 K. As the mantle cools, the steady-state water partitioning moves towards the bottom right in Fig.~\ref{fig:steadystate_temps}, sequestering water in the mantle at the expense of surface water.

\citet{schaefer16} modelled atmosphere-interior exchange on a hot M-Earth, GJ 1132b, to determine atmospheric composition, beginning with a magma ocean and allowing solidification, and including loss to space, but the authors only simulate the first few 100 Myr. Our simulations begin after magma ocean solidification, once the steam atmosphere has mostly condensed onto the surface, and plate tectonics permit cycling. Nonetheless, our initial partitioning is qualitatively consistent with \citet{schaefer16}, with the majority of water on the surface.

We are concerned with the surface water as it directly impacts habitability. We define four surface water regimes: a Dune planet, $10^{-5}$ TO $ \leq W_{\mathrm{s}} < 10^{-2}$ TO \citep{abe11}; an Earth-like regime, $10^{-2}$ TO $ \leq W_{\mathrm{s}} < 10$ TO; and a waterworld, where the surface is completely inundated, $W_{\mathrm{s}} \geq 10$ TO \citep{abbot12}\footnote{The waterworld definition of \citet{wordsworth13} is similar --- all land is covered by water, but this does not completely inhibit degassing from the interior. We assume this as well for M-Earths in the waterworld regime.}. We designate planets with $\lesssim 0.1$\% of the surface water of a Dune planet as desiccated. This is the amount of water currently in the atmosphere of Earth ($1.29 \times 10^{16}$ kg, or ${\sim}10^{-5}$ TO; \citealt{gleick93}); this amount of water is similar to Lake Superior. If precipitated onto the surface, it would produce a global ocean of depth ${\sim}2.5$ cm \citep{graham10}. While Dune planets, Earth-like planets, and waterworlds all have at least some liquid surface water and are thus habitable, only Earth-like planets are likely to have a silicate weathering thermostat.

If surface desiccation occurs, then regassing stops. Degassing will continue if the mantle is still warm; once the degassing rate surpasses the loss rate, surface water will increase. The surface is rehydrated when it exceeds our desiccation limit of ${\sim}10^{-5}$ TO.

\subsection{Individual Cycling Results}

\begin{figure*}
    \centering
    \includegraphics[width=0.45\textwidth]{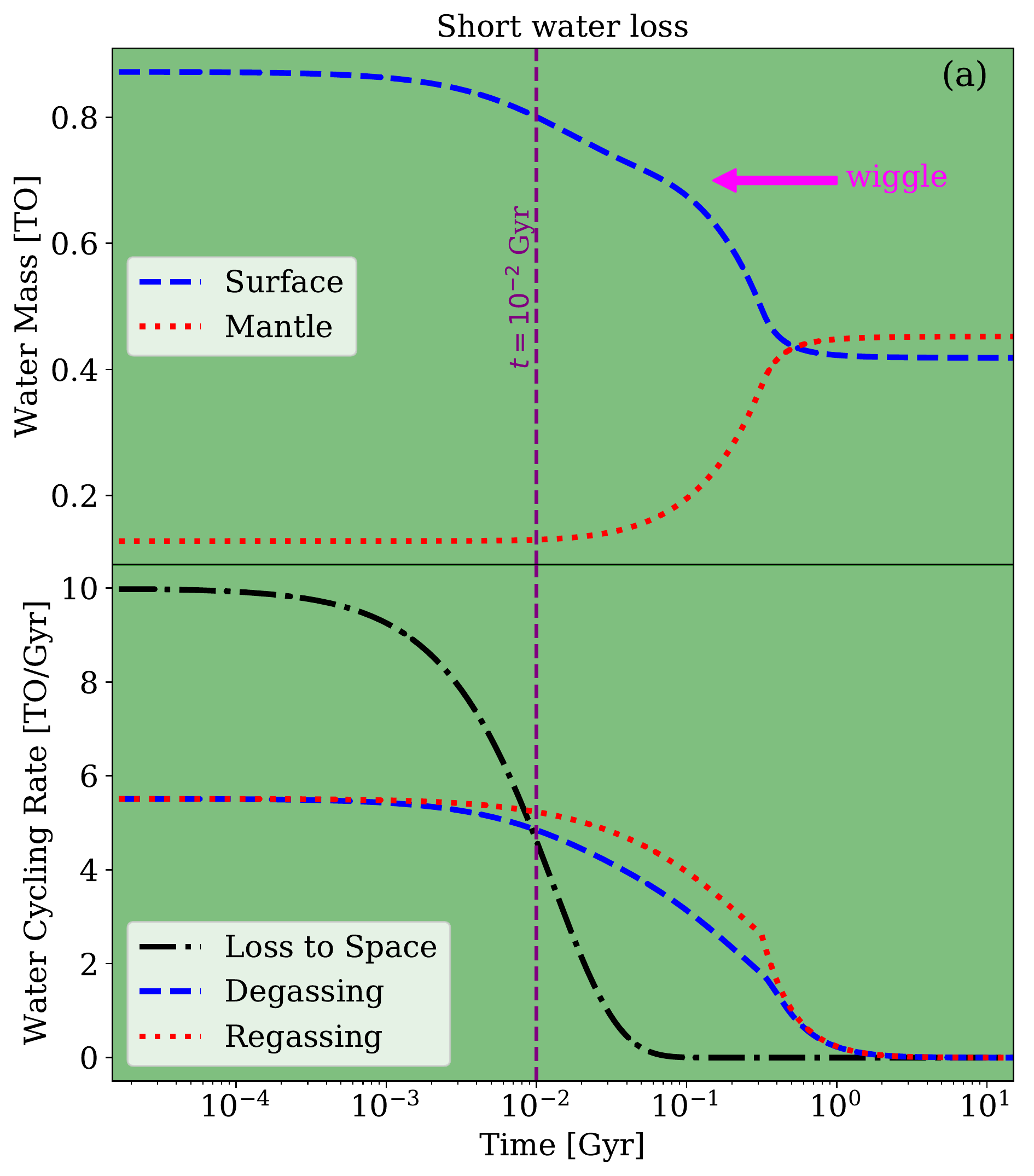}
    \includegraphics[width=0.45\textwidth]{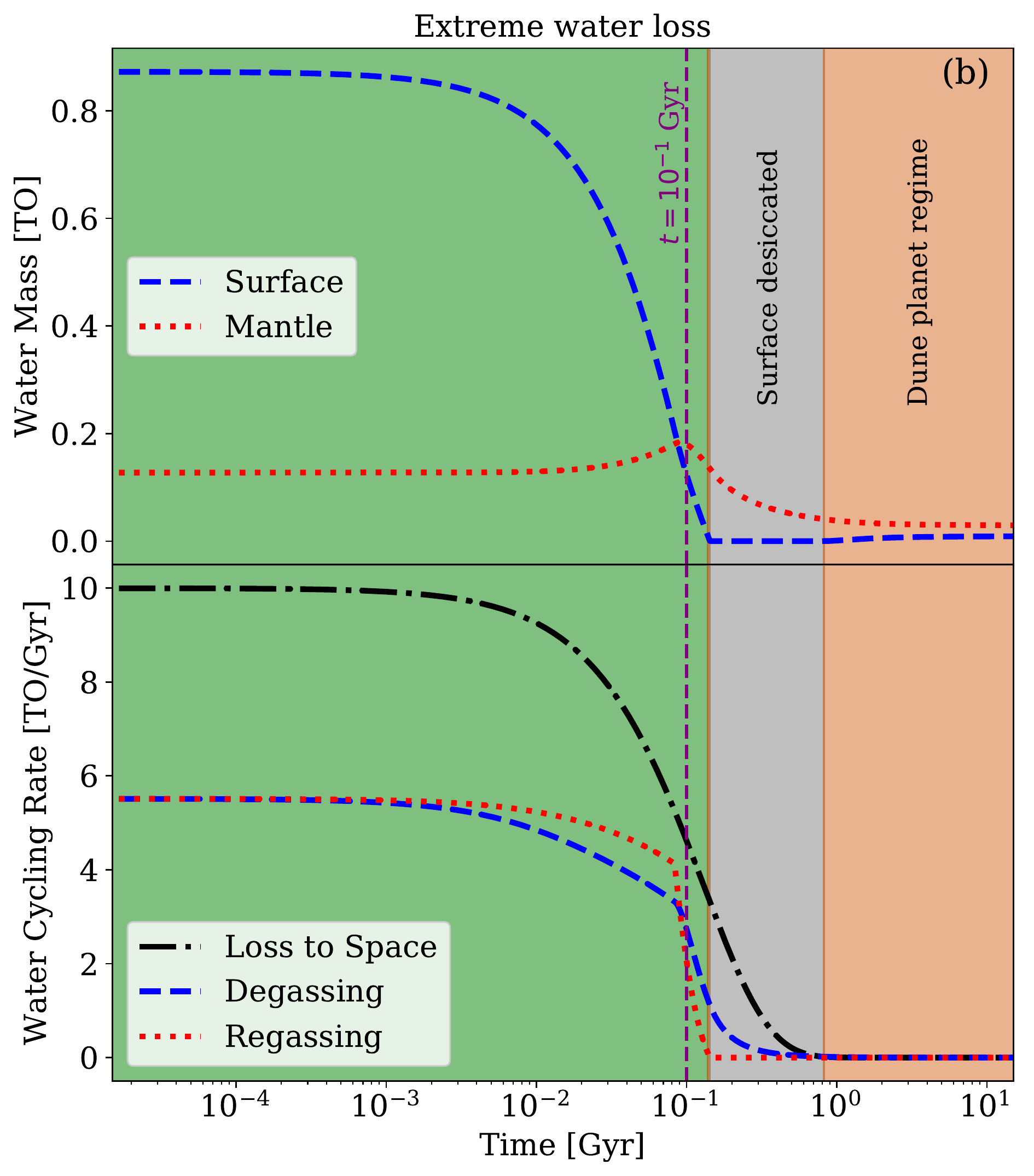}
    \caption{(a) Water cycling with short-lived loss of water to space. The top panel shows water partitioning between mantle, $W_{\mathrm{m}}$, and surface, $W_{\mathrm{s}}$, over time, while the bottom panel shows the evolution of the degassing, $w_{\uparrow}$, regassing, $w_{\downarrow}$, and loss, $w_{\mathrm{loss}}$, rates. Since we begin the simulation at steady-state partitioning, $w_{\uparrow,0} \simeq w_{\downarrow,0}$. The surface reservoir (top) is directly affected by loss to space; the loss rate is initially much higher than the cycling rates (bottom). Around $t = \tau_{\mathrm{loss}} = 10^{-2}$ Gyr, the cycling rates surpass the loss rate, causing the wiggle in $W_{\mathrm{s}}$ as cycling and loss compete to affect the surface water. Since regassing exceeds degassing during this time, some surface water is lost to space, while some is sequestered in the mantle. Eventually, the loss diminishes sufficiently to allow for a new steady state with a smaller total water inventory and cooler mantle temperature. The steady-state conditions will persist until the mantle cools below the solidus and degassing stops, which does not occur by 15 Gyr in this simulation. Despite the initial effect of water loss to space, the planet remains Earth-like throughout the simulation (as indicated by the green shaded region). (b) Extreme water loss, where the loss timescale is now $10\times$ longer than Fig.~\ref{fig:hyb_exploss}(a). Note that the plotted loss rate, $w_{\mathrm{loss}}$, in the bottom panel is an upper limit on the actual water lost, which is limited by the amount of water on the surface, $W_{\mathrm{s}}$. The loss of water to space causes rapid reduction of surface water, $W_{\mathrm{s}}$ (top); the planet briefly exists in the Dune planet regime (thin left brown region), but regassing, $w_{\downarrow}$, quickly approaches zero (bottom) as the remaining surface water is lost, approaching desiccation just after $t = \tau_{\mathrm{loss}} = 10^{-1}$ Gyr. During the grey region, the surface briefly becomes desiccated by loss (i.e., $W_{\mathrm{s}} \rightarrow 0$), which completely stops regassing, so cycling only occurs in one direction; water degassed after this time is immediately lost to space since the loss rate, $w_{\mathrm{loss}}$, still exceeds degassing, $w_{\uparrow}$. The degassing rate eventually surpasses the loss rate, and the surface is able to recover into the Dune planet regime (right brown region) before $t = 1$ Gyr. Since $\tau_{\mathrm{loss}}$ is $10\times$ longer than in Fig.~\ref{fig:hyb_exploss}(a), there is significantly less total water present on the planet by 15 Gyr; since the mantle remains warm and cycling continues, however, a new steady state is again approached.}
\label{fig:hyb_exploss}
\end{figure*}

We first show the time-dependent cycling and loss results for two simulations from the parameter exploration (Fig.~\ref{fig:hyb_exploss}). Both begin with $W_{\mathrm{m},0}+W_{\mathrm{s},0}$ = 1 TO of water, and loss factor $\phi_{\mathrm{loss}} = 10$ TO/Gyr. The ``short loss'' simulation uses a loss timescale of $\tau_{\mathrm{loss}} = 10^{-2}$ Gyr, while the ``extreme loss'' uses a longer loss timescale, $\tau_{\mathrm{loss}} = 10^{-1}$ Gyr. Since loss to space is occurring over a longer period for the latter, we expect a stronger reduction in surface water. 

Water cycling with short loss is shown in Fig.~\ref{fig:hyb_exploss}(a). Degassing, $w_{\uparrow}$, and regassing, $w_{\downarrow}$, are initially equal since we begin our cycling simulation from steady-state water partitioning. As a result, there is much more water on the surface, $W_{\mathrm{s}}$, than in the mantle, $W_{\mathrm{m}}$ (top panel). The loss rate to space, $w_{\mathrm{loss}}$, is initially higher than the cycling rates.

Loss to space initially dominates, reducing the surface water inventory, which reduces the amount available to sequester back into the mantle at late times. The cycling rates surpass the loss rate at $t \approx \tau_{\mathrm{loss}} = 10^{-2}$ Gyr, and as loss slows, the water partitioning seeks a new steady state for the current total water inventory and cooler mantle temperature, $T$.

The results for the extreme water loss simulation are shown in Fig.~\ref{fig:hyb_exploss}(b). The cycling begins similarly to the previous simulation, but the surface is rapidly desiccated (grey region). Degassing still provides water to the surface, where some is lost to space and a small amount is regassed, gradually reducing the mantle water inventory, while the surface water complement approaches zero.

Eventually, degassing from the mantle surpasses loss to space, and the surface recovers enough water to once again become a Dune planet (brown region). Even with continued cycling, there is not enough total water remaining on the planet to recover Earth-like surface conditions. Nonetheless, Fig.~\ref{fig:hyb_exploss}(b) demonstrates that water sequestered in the mantle can rehydrate the surface once loss to space diminishes.

\subsection{Parameter Exploration}

\begin{figure*}
\centering
\includegraphics[width=0.9\textwidth]{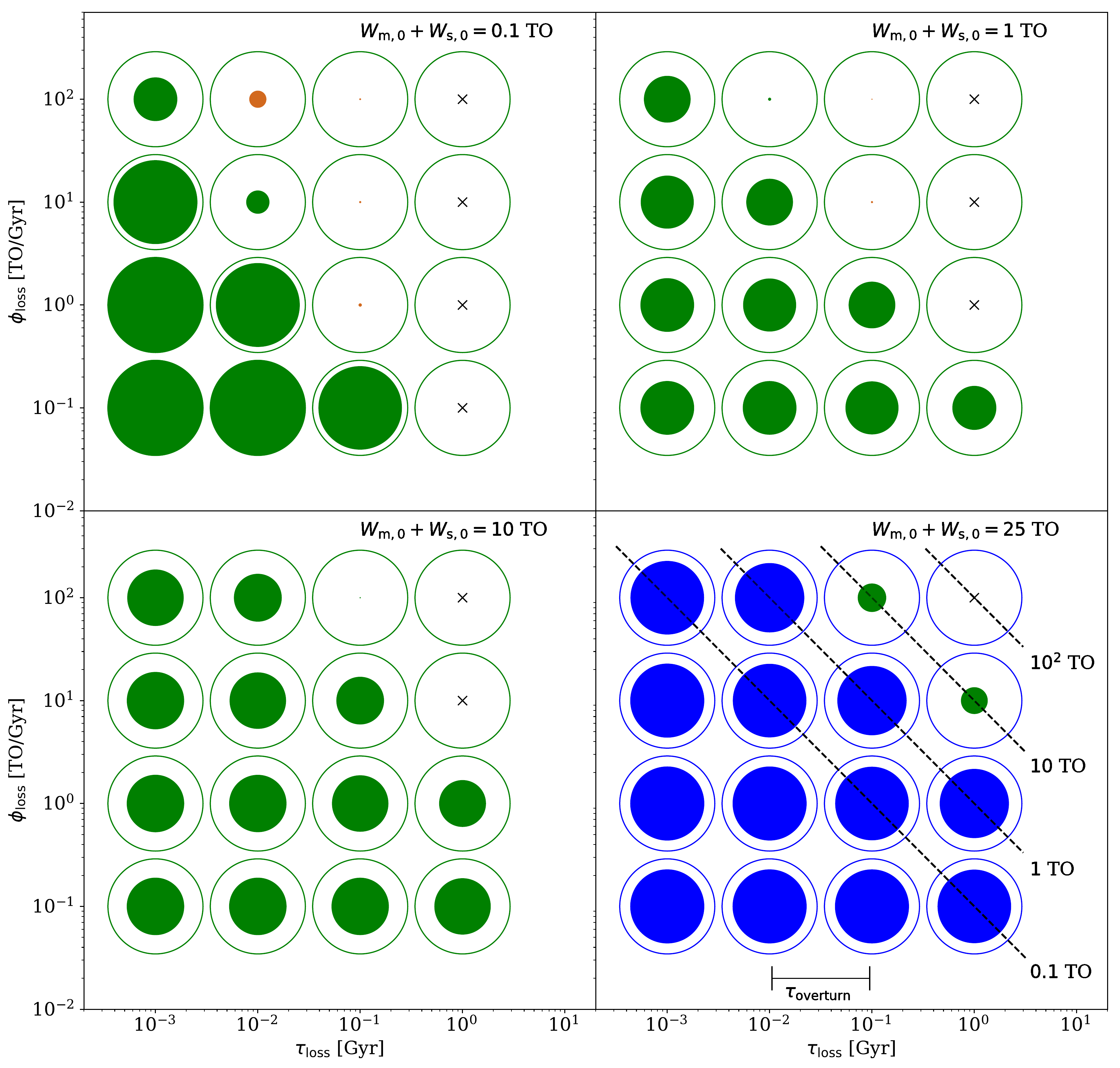}
\caption{\label{fig:Wloss_tloss_water} Evolution of surface water, $W_{\mathrm{s}}$, for different loss factors, $\phi_{\mathrm{loss}}$, and loss timescales, $\tau_{\mathrm{loss}}$.  The maximum amount of water a planet could lose is $\phi_{\mathrm{loss}} \times \tau_{\mathrm{loss}}$, as indicated in the bottom-right panel; diagonals correspond to simulations with equal $\phi_{\mathrm{loss}} \tau_{\mathrm{loss}}$. The open circles represent the amount of initial surface water, and the filled circles the surface water after 15 Gyr of cycling and loss to space. Colours indicate the surface water regime: waterworlds are blue, Earth-like planets are green, Dune planets are brown, and a desiccated surface is indicated by a black \textsf{x}. The initial total water inventories, $W_{\mathrm{m},0} + W_{\mathrm{s},0}$, are shown in the upper right of each panel, and filled circles are scaled based on the initial surface water of the open circles to visualize water loss. Planets subjected to water losses greater than their initial inventory, $\phi_{\mathrm{loss}} \tau_{\mathrm{loss}} \geq W_{\mathrm{m},0} + W_{\mathrm{s},0}$, would naively be expected to end up desiccated. Water sequestration in the mantle changes the picture dramatically, halving the simulations ending in desiccation. The approximate range of mantle overturn timescale, $\tau_{\mathrm{overturn}}$, is indicated in the bottom-right panel; mantle overturn is faster at early times and slows as the mantle cools. Planets are able to evolve between surface water regimes (e.g., Earth-like to Dune planet, or waterworld to Earth-like), but also able to recover water on a desiccated surface at later times by degassing water sequestered in the mantle.}
\end{figure*}

We now perform an exploration of the parameter space in Table \ref{tab:param_space}. We focus on the final surface water inventories to determine what surface conditions to expect after 15 Gyr of water cycling and loss to space. Our results for the 64 simulations are illustrated in Fig.~\ref{fig:Wloss_tloss_water}. We choose initial water inventories up to $W_{\mathrm{m},0}+W_{\mathrm{s},0}=25$ TO, far below the high-pressure ice limit of $W_{\mathrm{s,max}} = 100$ TO \citep{nakayama19}, to allow plate tectonic-driven cycling to continue uninhibited.

As shown in Fig.~\ref{fig:Wloss_tloss_water}, planets can evolve between surface water regimes. Certain rates of water loss to space cause waterworlds (blue) to lose sufficient water to expose continents and become Earth-like (similar to the ``waterworld self-arrest'' of \citealt{abbot12}), or Earth-like planets (green) to become dry Dune planets (brown) with little surface water, or even develop a completely desiccated, uninhabitable surface.

Water loss is limited by the amount of water on the surface (top left panel, Fig.~\ref{fig:Wloss_tloss_water}). A loss rate of 10 TO/Gyr and a loss timescale of 0.1 Gyr might in principle desiccate a planet with a 0.1 TO inventory, but since the loss predominantly happens early in the evolution, it only removes the surface water present at that time. 

Fig.~\ref{fig:Wloss_tloss_water} only shows the initial and final surface water contents, but we check for mid-simulation desiccation, as shown in Fig.~\ref{fig:hyb_exploss}(b). Ten of the 64 simulated planets recover from desiccation into either the Dune or Earth-like regime. This further supports our mechanism of sequestering water in the mantle and degassing it once atmospheric loss has decreased appreciably to restore habitable surface conditions.

\section{Discussion \& Conclusions}\label{sec:discussion}

Our simulations show that sequestering water in the mantle and subsequent degassing enhances the likelihood of habitable M-Earths in the face of atmospheric loss, provided they have an Earth-like deep-water cycle.

\subsection{Model Timescales}

There are three relevant timescales in our model that permit a more thorough interpretation of our results (Figs.~\ref{fig:hyb_exploss} \& ~\ref{fig:Wloss_tloss_water}). These timescales are the time to reach steady state, $\tau_{\mathrm{ss}}$, the loss timescale, $\tau_{\mathrm{loss}}$, and the mantle overturn timescale, $\tau_{\mathrm{overturn}}$.

The surface water content and mantle temperature in our model change with time, not only due to loss but also water degassed and regassed from and to the mantle, respectively. The planet will therefore be approaching a changing steady state with time (Fig.~\ref{fig:steadystate_temps}). This steady state will not be reached until loss diminishes and mantle cooling slows at late times, allowing degassing and regassing rates to equilibrate. Indeed, we only see steady state achieved late in our simulations, and only as long as the mantle remains warm.

If the loss timescale is much longer than the mantle overturn timescale, $\tau_{\mathrm{loss}} \gg \tau_{\mathrm{overturn}}$, the water lost to space will be roughly $\phi_{\mathrm{loss}} \tau_{\mathrm{loss}}$, provided there is sufficient total water on the planet. This explains the different results for the same $\phi_{\mathrm{loss}} \tau_{\mathrm{loss}}$ in Fig.~\ref{fig:Wloss_tloss_water}. Since the time to reach steady-state is closely related to the mantle overturn timescale, $\tau_{\mathrm{loss}} \gg \tau_{\mathrm{overturn}}$ also means that the planet is always at or near a steady state (equal degassing and regassing), but that steady state is a moving target due to atmospheric loss.

If $\tau_{\mathrm{loss}} \ll \tau_{\mathrm{overturn}}$, however, the total water lost is now limited by the initial surface water on the planet, $W_{\mathrm{s},0}$. Since most loss happens early on and the loss rate diminishes with time, the surface can eventually be rehydrated, provided there is sufficient water sequestered within the mantle. This explains the similar results seen in each panel of Fig.~\ref{fig:Wloss_tloss_water}, on the left and bottom-left. The greater a planet's initial water inventory, the farther towards the upper-right corner this plateau extends.

In summary, the total amount of water lost, $W_{\mathrm{lost}}$, is,
\begin{equation}
    W_{\mathrm{lost}} = \begin{cases}
    \min[\phi_{\mathrm{loss}} \tau_{\mathrm{loss}}, ~W_{\mathrm{s},0}+W_{\mathrm{m},0}] & \tau_{\mathrm{loss}} \gg \tau_{\mathrm{overturn}} \\
    \min[\phi_{\mathrm{loss}} \tau_{\mathrm{loss}}, ~W_{\mathrm{s},0}] & \tau_{\mathrm{loss}} \ll \tau_{\mathrm{overturn}}.
    \end{cases}
\end{equation}

\subsection{Thermal Evolution \& Tectonic Mode}

The model can approach a steady state once loss has diminished significantly, as long as the mantle remains above the solidus temperature. Once the mantle cools below the solidus temperature, degassing stops due to the absence of melt below mid-ocean ridges. This leads to regassing-dominated evolution, eventually trapping all water in the mantle \citep{schaefer15}. 

It has been postulated, however, that when the mantle cools below the solidus temperature or becomes desiccated, convection may stop, along with plate tectonics, transitioning to a ``stagnant lid'' regime \citep{noack14, lenardic18}. As noted by \citet{schaefer15}, transitioning to a stagnant lid would stop both degassing and regassing, preserving the water inventories in surface and mantle reservoirs at that time. A stagnant lid would greatly affect our cycling parameterizations, but volatiles can still be cycled in a stagnant-lid regime, albeit at a much slower rate \citep{honing19}. We leave this complication for future work; however, since our current simulations sometimes regas all water into the mantle, presumably accounting for a stagnant lid would merely result in more surface water at late times.

\subsection{Observational Prospects}

Observationally characterizing M-Earth atmospheres in the near future is viable \citep{cowan15, shields16, gillon20}, but direct detection of surface water on an exoplanet is probably still many years away \citep{cowan09, robinson10, lustig18}. To zeroth-order, our conclusions support continued observations of M dwarf systems in the search for habitability. Our results will be useful in interpreting observations, allowing inference of the cycling \& loss history of M-Earths based on, e.g., the presence of H\textsubscript{2}O in transit spectra. Connecting surface water to climate, atmospheric structure, and transit spectroscopy will be the subject of a future study. Increasing the fidelity of our M-Earth water cycling \& loss model will narrow the gap between predictions and observations.

The key variables in our model include the initial water inventory, the initial water partitioning, the mantle overturn timescale, the loss rate and the loss timescale. The initial water inventory may be difficult to determine due to the stochastic nature of volatile delivery during planet formation \citep{raymond04, raymond09}; however, studies of volatiles in protoplanetary disks (e.g., using ALMA; \citealt{harsono20}; \citealt{loomis20}) and studies of polluted white dwarfs (e.g., \citealt{farihi16}; \citealt{veras17a, veras17b}; \citealt{doyle19}) may provide constraints. The geophysical processes that determine both the planetary water partitioning and mantle overturn timescale in our M-Earth model are based on present-day Earth. Determining the tectonic mode of an observed exoplanet will be difficult in the near future, but in principle, may be possible with LUVOIR (e.g., \citealt{cowan09}); nonetheless, modelling can allow exploration of the potential geophysics on distant planets. Many uncertainties in our model arise due to our treatment of stellar evolution, but we may be able to constrain the loss rate (i.e., the loss factor and timescale) through a combination of M dwarf observations and modelling to better represent the governing loss processes on an M-Earth.

\section*{Acknowledgements}

We thank the anonymous referee for a beneficial referee report that strengthened this manuscript. We thank Tad Komacek for providing his hybrid model and for valuable correspondence while re-coding the model. We also acknowledge thesis committee members Yajing Liu, Vincent van Hinsberg, Don Baker, Galen Halverson, and Timothy Merlis, as well as insightful conversations with Christie Rowe and Mark Jellinek. We thank Dylan Keating and Lisa Dang for comments on a draft of this manuscript. K.M. acknowledges support from a McGill University Dr. Richard H. Tomlinson Doctoral Fellowship, and from the Natural Sciences and Engineering Research Council of Canada (NSERC) Postgraduate Scholarships-Doctoral Fellowship. 

\section*{Data Availability}

The model and data presented within this manuscript are available from the corresponding author at reasonable request.

\bibliographystyle{mnras}

\appendix

\section{Thermal Evolution Equations}\label{sec:thermaleqns}

The thermal evolution of the mantle in our model, which incorporates parameterized convection, is a simplified version of the thermal evolution presented within \citet{schaefer15}, itself based on the model of \citet{sandu11}. For simplicity, and to reproduce the low-viscosity model which reaches an analytic steady state (the goal of \citealt{komacek16}), the convection is constrained to a boundary layer in the upper mantle.

The evolution of the mantle temperature, $T$, with time, $t$, in our model is dependent on mantle water mass fraction $x = W_{\mathrm{m}}/ M_{\mathrm{m}}$, where $M_{\mathrm{m}}$ is the mass of the mantle. The thermal evolution equation is: 
\begin{equation}\label{eqn:dT/dt}
\begin{split}
     \rho_{\mathrm{m}} c_{\mathrm{p}} \frac{dT}{dt} & = Q(t) - \frac{A}{V} \frac{k \Delta T}{h} \left(\frac{\mathrm{Ra}}{\mathrm{Ra}_{\mathrm{crit}}} \right)^\beta \\
     & = Q(t) - \frac{k A \Delta T}{h V} \left(\frac{\alpha \rho_{\mathrm{m}} g h^3 \Delta T}{\mathrm{Ra}_{\mathrm{crit}} \kappa \eta(T,x)} \right)^\beta.
\end{split}
\end{equation}
The value of $\beta = 0.3$ in this equation was determined empirically \citep{mcgovern89}. The density of the upper mantle is $\rho_{\mathrm{m}}$, and $c_{\mathrm{p}}$ is the mantle's specific heat capacity. The heating rate from radionuclides is $Q(t) = Q_0 \exp^{-t/\tau_{\mathrm{decay}}}$. The decay timescale $\tau_{\mathrm{decay}} = 2$ Gyr was nominally chosen by \citet{komacek16}, based on the abundance and half-lives of radiogenic elements in the Earth's mantle from \citet{turcotte02}. For reference, this value falls between the half-lives of \textsuperscript{40}K (1.3 Gyr) and \textsuperscript{238}U (4.5 Gyr). The thermal conductivity of the upper mantle is $k$, and $\Delta T = T - T_{\mathrm{s}}$ is the temperature contrast across the boundary layer, with $T_{\mathrm{s}}$ the surface temperature.

The scaling laws of terrestrial planets from \citet{valencia06} allow us to calculate the planet's mantle thickness, $h$, the planet's surface area, $A$, the mantle volume, $V$, and surface gravity, $g$. Planetary radius, $R$, and core radius, $R_{\mathrm{c}}$, are related to planetary mass, $M$, by
\begin{equation}\label{eqn:R}
    R = R_{\oplus} \left(\frac{M}{M_{\oplus}} \right)^p,
\end{equation}

\begin{equation}\label{eqn:R_c}
    R_{\mathrm{c}} = c R_{\oplus} \left(\frac{M}{M_{\oplus}} \right)^{p_{\mathrm{c}}},
\end{equation}
where $p=0.27$, $c=0.547$, and $p_{\mathrm{c}}=0.25$. The remaining planet and mantle properties are:
\begin{equation}\label{eqn:h}
    h = R - R_{\mathrm{c}},
\end{equation}

\begin{equation}\label{eqn:A}
    A = 4 \pi R^2,
\end{equation}

\begin{equation}\label{eqn:V}
    V = \frac{4 \pi}{3} (R^3 - R_{\mathrm{c}}^3),
\end{equation}

\begin{equation}\label{eqn:g}
    g = \frac{G M}{R^2}.
\end{equation}

The Rayleigh number of the mantle, $\mathrm{Ra} = \alpha \rho_{\mathrm{m}} g h^3 \Delta T/ \kappa \eta(T,x)$, can be calculated using the upper mantle density, $\rho_{\mathrm{m}}$, mantle thickness, $h$, and temperature contrast, $\Delta T$, along with the characteristic thermal expansivity, $\alpha$, the planet's gravity, $g$, and the thermal diffusivity of the boundary layer, $\kappa$. The critical Rayleigh number for convection to occur in the upper mantle is $\mathrm{Ra}_{\mathrm{crit}} = 1100$ \citep{mcgovern89}. Due to the dependence of temperature $T$ on mantle water mass fraction $x$, the thermal evolution and cycling equations are integrated simultaneously.

For the mantle viscosity, $\eta(T,x)$, we use the same parameterization as \citet{komacek16}, which in turn is a simplified version of that from the models of \citet{sandu11} and \citet{schaefer15} (i.e., without the pressure-dependence, since we are restricted to the upper mantle):
\begin{equation}\label{eqn:eta}
    \eta \approx \eta_0 f_{\mathrm{w}}^{-r} \exp \left[\frac{E_\mathrm{a}}{R_{\mathrm{gas}}} \left( \frac{1}{T} - \frac{1}{T_{\mathrm{ref}}} \right) \right].
\end{equation}
Here, $\eta_0$ is the viscosity scale (chosen so that $\eta(x=x_{\oplus}, T=T_{\mathrm{ref}}) = 10^{21}$ Pa$\cdot$s to reproduce the viscosities of Earth's mantle), $E_{\mathrm{a}}$ is the activation energy, $R_{\mathrm{gas}}$ is the universal gas constant, $T_{\mathrm{ref}}=1600$ K is the reference mantle temperature, $f_{\mathrm{w}}$ is water fugacity (see Eqn.~\ref{eqn:f_w} below), and $r=1$ is the nominal value chosen by \citet{schaefer15}, based on measurements of wet olivine diffusion.

The water fugacity, $f_{\mathrm{w}}$, can be calculated using experimental data from \citet{li08},
\begin{equation}\label{eqn:f_w}
\begin{split}
    \ln f_{\mathrm{w}} & = c_0 + c_1 \ln \left( \frac{B x \mu_{\mathrm{oliv}}/\mu_{\mathrm{w}}}{1 - x \mu_{\mathrm{oliv}}/\mu_{\mathrm{w}}} \right) + c_2 \ln^2 \left( \frac{B x \mu_{\mathrm{oliv}}/\mu_{\mathrm{w}}}{1 - x \mu_{\mathrm{oliv}}/\mu_{\mathrm{w}}} \right) \\
    & + c_3 \ln^3 \left( \frac{B x \mu_{\mathrm{oliv}}/\mu_{\mathrm{w}}}{1 - x \mu_{\mathrm{oliv}}/\mu_{\mathrm{w}}} \right),
\end{split}
\end{equation}
where $c_0 = -7.9859$, $c_1 = 4.3559$, $c_2 = -0.5742$, $c_3 = 0.0337$, $B=2 \times 10^6$ (which converts to number concentration of H atoms per $10^6$ Si atoms), $\mu_{\mathrm{oliv}}$ is the molecular weight of olivine, and $\mu_{\mathrm{w}}$ is the molecular weight of water.

\section{Model Improvements \& Constraints}\label{sec:improvements}

Our improvements to the hybrid model of \citet{komacek16} include restoring the degassing limit from \citet{cowan14}, and capacity limits for mantle water (to account for a saturated mantle, $W_{\mathrm{m}} \leq 12$ TO; \citealt{hauri06, cowan14}) and surface water ($W_{\mathrm{s}} \leq 100$ TO, above which high-pressure ices will form at the ocean floor and significantly alter or hinder the degassing/regassing rates; \citealt{nakayama19}).

The addition of a simultaneous loss term in the cycling equations brings the model closer to predictions for XUV-driven water loss to space on M-Earths (e.g. \citealt{luger15}). Our current loss factors, $\phi_{\mathrm{loss}}$, and timescales, $\tau_{\mathrm{loss}}$, allow for a phenomenological exploration of parameter space to show the effect of water loss rather than being directly calculated from, for example, stellar evolution models of XUV flux from M dwarfs (e.g., \citealt{ribas05}). 

While coupled integrations of the thermal evolution in Eqn.~\ref{eqn:dT/dt} and the cycling of Eqns.~\ref{eqn:dx/dt_hyb} and \ref{eqn:ds/dt_hyb} can be performed using the \texttt{scipy.integrate} package in Python, restrictions must be placed to ensure we do not obtain meaningless, unphysical results. Our hybrid model is robust to either or both reservoirs going to zero, an improvement over \citet{komacek16}. At each timestep, surface and mantle water inventories are checked, and the cycling equations are adjusted accordingly:
\begin{enumerate}
    \item If mantle water mass $W_{\mathrm{m}}$ and surface water mass $W_{\mathrm{s}}$ are both greater than zero, then normal cycling and loss occurs, and the cycling equations are integrated as they appear in \S\ref{sec:equations}.
    
    \item If $W_{\mathrm{m}} = 0$ but $W_{\mathrm{s}} > 0$, the degassing rate, $w_{\uparrow}$, is set to zero (i.e., it shuts off since there is no water in the mantle) in Eqns.~\ref{eqn:dx/dt_hyb} and \ref{eqn:ds/dt_hyb}, and the integration is performed.
    
    \item If $W_{\mathrm{m}} > 0$ but $W_{\mathrm{s}} = 0$, the regassing rate, $w_{\downarrow}$, and loss rate, $w_{\mathrm{loss}}$, are set to zero (since there is no water on the surface) in Eqns.~\ref{eqn:dx/dt_hyb} and \ref{eqn:ds/dt_hyb}, and the integration is performed. We also set the fraction of water in melt that is degassed $f_{\mathrm{degas}}(P)=1$, due to its piecewise definition. We do this because the $P$-dependent degassing rate depends on the overlying surface water, $W_{\mathrm{s}}$; if there is no water on the surface, degassing should neither go to zero and shut off (or else water would stay in the mantle indefinitely), nor go to infinity (or all water would be instantaneously degassed from the mantle).
    
    \item If both $W_{\mathrm{m}} = 0$ and $W_{\mathrm{s}} = 0$, the degassing, regassing, and loss rates are all set to zero, since the planet is completely desiccated. The integration continues so we can observe the thermal evolution of the mantle (i.e., its cooling with time), but there is no cycling or loss since there is no more water present on the surface or within the mantle of the planet.
\end{enumerate}

Our piecewise definition of loss in Eqn.~\ref{eqn:ds/dt_hyb} ensures that the amount of surface water that is regassed and lost at a given timestep does not exceed the amount present on the surface, and the regassing/loss rates are adjusted accordingly based on the surface water mass, $W_{\mathrm{s}}$.

Although it is a result we have yet to encounter, to account for complete mantle desiccation (a scenario proposed in the literature; see, e.g., \citealt{hamano13}), we choose a minimum water fugacity to avoid $f_{\mathrm{w}} \rightarrow 0$ and mantle viscosity $\eta \rightarrow \infty$. We can then define a piecewise mantle water fugacity, $f_{\mathrm{w,eff}}$, represented by the equation,
\begin{equation}\label{f_w_limit}
    f_{\mathrm{w,eff}} = \max [10^{-5} f_{\mathrm{w}}(\tilde{x}=1), ~f_{\mathrm{w}}]
\end{equation}
where $f_{\mathrm{w}}(\tilde{x}=1)$ is used to define the non-dimensional fugacity within our model code, $\tilde{f}_\mathrm{w} = f_\mathrm{w} / f_{\mathrm{w}}(\tilde{x}=1)$. This definition requires the non-dimensional water mass fraction $\tilde{x} = x f_\mathrm{m} / (\omega_0 \tilde{f}_\mathrm{b})$, where $f_{\mathrm{m}}$ is the mantle fraction, $\omega_0$ is the surface water mass fraction of Earth, and $\tilde{f}_\mathrm{b} = f_\mathrm{b} / f_{\mathrm{b},\oplus} = 1.3$ is the non-dimensional ocean basin covering fraction, with $f_{\mathrm{b},\oplus}=0.7$ and $f_\mathrm{b} = 0.9$ (i.e., 90\% of planet covered in water). The value for $f_{\mathrm{b}}$ was chosen by \citet{cowan14}, which we also optimistically adopt for an Earth-like planet. 

The minimum value, $f_{\mathrm{w,eff}} = 10^{-5} f_\mathrm{w}(\tilde{x}=1)$, assumes that even in the case of a completely desiccated mantle, there will be a small amount of water trapped in the minerals (e.g., within the transition zone; \citealt{hirschmann06}). This allows the mantle to continue convecting and our plate-tectonics-dependent cycling to proceed. Note that, throughout the thermal evolution equations presented above, $f_{\mathrm{w}}$ is used in place of $f_{\mathrm{w,eff}}$ for consistency with the literature.

Finally, we note that the water fugacity was calculated incorrectly in many places in the original hybrid cycling model due to a misplaced bracket (Komacek 2019, priv.~comm.). While this error does not significantly impact the final results, it does slightly change the time-dependent cycling results, and has been fixed in the model presented here.

\section{Model Parameters}\label{sec:params}

There are many variables throughout this paper. As such, we detail them all in Table \ref{tab:modelparams}, including their nominal values.

\begin{table*}
    \centering
    \begin{tabular}{c|c|c}
    \hline
        Name & Parameter & Value  \\
    \hline
        Mantle water mass$^{ab}$ & $W_{\mathrm{m}}$ & $W_{\mathrm{m},\oplus} = 2.36 \times 10^{21}$ kg $\approx 1.7$ TO \\
        Surface water mass$^{ab}$ & $W_{\mathrm{s}}$ & $W_{\mathrm{s},\oplus} \approx 1.4 \times 10^{21}$ kg $ = 1$ TO \\
        Mantle temperature$^a$ & $T$ & $T_{\mathrm{ref}} = 1600$ K \\
        Loss factor$^b$ & $\phi_{\mathrm{loss}}$ & $10$ TO Gyr$^{-1}$ \\
        Loss timescale$^b$ & $\tau_{\mathrm{loss}}$ & $10^8$ yr \\
        Mid-ocean ridge length & $L_{\mathrm{MOR}}$ & $L_{\mathrm{MOR},\oplus} = 60 \times 10^6$ m \\
        Spreading rate$^a$ & $S$ &  $S_{\oplus,\mathrm{avg}} \approx 0.1$ m yr$^{-1}$ \\
        Water mass fraction in hydrated crust & $x_{\mathrm{h}}$ & 0.05 \\
        Crust density & $\rho_{\mathrm{c}}$ & $3.0 \times 10^3$ kg m$^{-3}$ \\
        Regassing efficiency & $\chi_{\mathrm{r}}$ & 0.03 \\
        Hydrated layer depth$^a$ & $d_{\mathrm{h}}$ & $d_{\mathrm{h},\oplus} = 3.0 \times 10^3$ m \\
        Mantle water mass fraction & $x$ & $x_{\oplus} = 5.8 \times 10^{-4}$ \\
        Mantle density & $\rho_{\mathrm{m}}$ & $3.3 \times 10^3$ kg m$^{-3}$ \\
        Mid-ocean ridge melting depth & $d_{\mathrm{melt}}$ & $60 \times 10^3$ m \\
        Melt degassing efficiency of Earth & $f_{\mathrm{degas},\oplus}$ & 0.9 \\
        Seafloor pressure$^a$ & $P$ & $P_{\oplus} = 4 \times 10^7$ Pa \\
        Timestep & $\tau_{\mathrm{step}}$ & ${\sim}28700$ yr \\
        Mantle overturn timescale & $\tau_{\mathrm{overturn}}$ & $\tau_{\mathrm{overturn},\oplus} \approx 6 \times 10^6$ yr \\
        Mantle specific heat capacity & $c_{\mathrm{p}}$ & 1200 J kg$^{-1}$ K$^{-1}$ \\
        Radionuclide decay factor & $Q_0$ & $5 \times 10^{-8}$ J m$^{-3}$ s$^{-1}$ \\
        Decay timescale & $\tau_{\mathrm{decay}}$ & 2 Gyr \\
        Mantle thermal conductivity & $k$ & 4.2 W m$^{-1}$ K$^{-1}$ \\
        Surface temperature & $T_{\mathrm{s}}$ & 280 K \\
        Mantle critical Rayleigh number & $\mathrm{Ra}_{\mathrm{crit}}$ & 1100 \\
        Heat flux exponent & $\beta$ & 0.3 \\
        Mantle characteristic thermal expansivity & $\alpha$ & $2 \times 10^{-5}$ K$^{-1}$ \\
        Mantle thermal diffusivity & $\kappa$ & $10^{-6}$ m$^2$ s$^{-1}$ \\
        Planet radius & $R$ & $R_{\oplus} = 6.371 \times 10^6$ m \\
        Planet mass & $M$ & $M_{\oplus} = 5.972 \times 10^{24}$ kg \\
        Gravitational constant & $G$ & $6.67 \times 10^{-11}$ m$^3$ kg$^{-1}$ s$^{-2}$ \\
        Mantle viscosity$^a$ & $\eta$ & $\eta(x=x_{\oplus}, T=T_{\mathrm{ref}}) = 10^{21}$ Pa s\\
        Water fugacity$^a$ & $f_{\mathrm{w}}$ & $f_{\mathrm{w}}(x_{\oplus}) \approx 17 \times 10^3$ Pa \\
        Fugacity exponent & $r$ & 1 \\
        Activation energy & $E_{\mathrm{a}}$ & $335 \times 10^3$ J mol$^{-1}$ \\
        Universal gas constant & $R_{\mathrm{gas}}$ & 8.314 J mol$^{-1}$ K$^{-1}$ \\
        Molecular weight of olivine & $\mu_{\mathrm{oliv}}$ & 153.31 g mol$^{-1}$ \\
        Molecular weight of water & $\mu_{\mathrm{w}}$ & 18.02 g mol$^{-1}$ \\
        Planetary mantle fraction & $f_{\mathrm{m}}$ & 0.68 \\
        Ocean basin covering fraction & $f_{\mathrm{b}}$ & 0.9 \\
        Earth ocean basin covering fraction & $f_{\mathrm{b},\oplus}$ & 0.7 \\
    \hline
    \multicolumn{3}{c}{{\bf Notes.}} \\
    \multicolumn{3}{c}{$^a$: These parameters are calculated during our coupled thermal evolution and cycling \& loss integrations.} \\
    \multicolumn{3}{c}{$^b$: These parameters are varied during our parameter exploration.} \\
    \end{tabular}
    \caption{Parameters \& constants used in our M-Earth thermal evolution and cycling \& loss model. The corresponding equations from which they are taken appear throughout this paper.}
    \label{tab:modelparams}
\end{table*}

\bsp	
\label{lastpage}
\end{document}